\documentclass[preprint,11pt]{article}

\usepackage{bm}
\usepackage{epsfig}
\usepackage{color}
\usepackage{amssymb}
\usepackage{amsmath}

\newcommand {\lab}[1]{\label{eq:#1}}

\newcommand {\be}[1]{\begin{equation}{\lab{#1}}}
\newcommand {\ee}{\end{equation}}
\newcommand {\bea}{\begin{eqnarray}}
\newcommand {\eea}{\end{eqnarray}}

\begin{document}

\title{Extensive packet excitations in FPU and Toda lattices}

\author{
\textbf{Helen Christodoulidi$^{1}$ }\\
$^{1}$Research Center for Astronomy and Applied Mathematics, Academy of Athens, Athens, Greece}


\maketitle

\begin{abstract}
At low energies, the excitation of low frequency packets of normal modes in the Fermi--Pasta--Ulam (FPU) 
and in the Toda model leads to exponentially localized energy profiles which resemble staircases and are 
identified by a slope $\sigma$ that depends logarithmically on the specific energy $\varepsilon=E/N$. 
Such solutions are found to lie on stable lower dimensional tori, named $q$--tori.
At higher energies there is a sharp transition of the system's localization profile
to a straight--line one, determined by an $N$--dependent slope of the form $\sigma \sim (\varepsilon N)^{-d}$, $d>0$.
We find that the energy crossover $\varepsilon _c$ between the two energy regimes decays as $1/N$, 
which indicates that $q$--tori disappear in the thermodynamic limit. Furthermore, we focus on the 
times that such localization profiles are practically frozen and we find that these `stickiness times'
can rapidly and accurately distinguish between a power--law and a stretched exponential dependence in $1/\varepsilon $.
\end{abstract}

                             

\section{Introduction}
\label{intro}

One of the most classical open problems of statistical mechanics is the Fermi--Pasta--Ulam paradox, 
which owes its name to the numerical experiment of Fermi, Pasta and Ulam \cite{feretal1955} in 1954 for probing ergodicity 
in an one--dimensional chain of $N$ weakly nonlinearly coupled oscillators. 
Fermi expected that even a slight perturbation in a solvable system, 
such as the linearly coupled oscillators, will result to a fully ergodic behavior. 
The reach of thermal equilibrium can be implied by the energy equipartition among all the normal modes, 
and therefore, by exciting the first normal mode ergodicity could be straight--forwardly observed.  
Nevertheless, in the numerical simulations performed by M. Tsingou \cite{tsingou}  a fast energy 
transfer to few consecutive modes was observed, followed shortly afterwards, by an energy backflow 
to the first mode, which formed the so--called FPU recurrences. This recursive phenomenon 
indicates more a quasiperiodic behavior instead of the anticipated ergodic one. 
Further reviews and details on the FPU problem and on the progress which has been done over the 
last 50 years can be found in \cite{bergman,FPUbook}.

Deriving rigorous results which can explain the integrable--like behavior in the FPU system is a 
particularly difficult task. In \cite{Rink1,Rink2} the existence of full dimensional
KAM--tori was shown in the FPU--$\beta $ model. Nevertheless, the Diophantine type of nonresonance condition 
needed for KAM--tori implies that 
they vanish exponentially with $N$. On the other hand, some particular solutions show a slower 
decay to zero due to less strict conditions. For example, 
considering simple periodic orbits in the periodic FPU--$\beta $ model, as is the $\pi$--mode $q=N/2$,  
the stability threshold in terms of the specific energy vanishes like $1/N^2$ 
\cite{Budinsky,flach_pi,Ruffo_pi_mode,Antonopoulos}. 
Another type of periodic orbits, closely related to the original FPU initial condition,
are the $q$--breathers, derived by the Lyapunov continuation of single normal modes
\cite{Flach2005,flaetal2006,flaetal2006b,flapon2008,flaetal2007,kanetal2007,Flachtiz}. 
Such solutions are well known for their strong energy localization properties, even beyond
their stability threshold $\varepsilon _{br} \sim q_0^4/N^4$, where $q_0$ is the `seed' normal
mode \cite{flaetal2006}.

 {In \cite{tori1,tori2} has been found that packets of linear normal modes (phonons) 
give rise to solutions on stable low--dimensional tori in the FPU phase space, called $q$--tori, and} whose energy spectrum profile  
resembles a `staircase' in Fourier space, constituted by exponentially decaying groups of modes. These tori have an 
 {intensive} property, namely, the slope of the exponential energy profile does not depend on $N$, as long as we
consider {\it extensive packets} of normal modes. On the other hand, the persistence of the $q$--tori
in the thermodynamic limit was left open in \cite{tori1,tori2}.

In the present work we provide evidence that, despite their extensive properties, the $q$--tori vanish towards the thermodynamic limit
($N\rightarrow \infty $ with constant energy per particle $\varepsilon =E/N$) 
like $\varepsilon _c \propto  1/N$. Nevertheless, the energy localization phenomenon persists
beyond $q$--tori's energy range.
In this higher energy regime, 
we find a different localization law, namely the energy spectrum has a straight--line profile 
with an $N$--dependent slope of the form $\sigma \sim (\varepsilon N)^{-d}$ for $d\approx 0.3$. 
The stability and localization of FPU--trajectories for $\varepsilon >\varepsilon _c $ 
cannot be derived 
from the coupled harmonic oscillators. The higher order tangency between the Toda system and
FPU implies that Toda is the most appropriate discrete integrable reference model to the FPU
\cite{ferguson,dresden,benettin2}. In other words, we conjecture that the integrable 
counterparts for long--term stable FPU--trajectories are not anymore the normal modes but the
Toda tori, therefore any construction based on the harmonic frequency approximation fails in this 
second regime and a set of non--commensurable Toda frequencies is what should be looked for.

Finally,  {in this work we examine the {\it stickiness times} or stability times, namely, 
the times which solutions stay close to their integrable counterparts. 
Up to our knowledge, the stickiness times have not been considered for models like FPU, in contrast to
the equipartition times, which have been posed already at the birth of the FPU problem.
A central question concerning the equipartition times is to distinguish between a power--law dependence
in  $1/\varepsilon$ and a stretched exponential, for the reason that exponentially--long times imply long--term 
stability with reference to Nekhoroshev theory \cite{nek}.
Many works \cite{benettin1,benettin2,Casetti,dresden} find that $T^{eq} \sim \varepsilon ^{-a}$, $a>0$
which however, becomes a stretched exponential $T^{eq} \sim  \exp (1/{\varepsilon }^{\delta } )$, 
$\delta >0$ below a certain energy threshold \cite{benettin2,beretal2004,beretal2005}. 
We find and report that it is faster to consider the stickiness times, which can accurately and reliably detect 
Nekhoroshev--regimes, where a direct implementation of the KAM theorem is almost impossible. }

\section{Packet excitations in FPU and Toda} \label{single}
The dynamics of the FPU model is 
described by the Hamiltonian function\footnote{In reality, this system is the `FPU--$\alpha$ model', 
which for simplicity we will call here it solely FPU.}: 
\begin{eqnarray}\label{fpuham} 
H={1\over 2}\sum_{k=1}^N y_k^2 + {  1   \over
2}\sum_{k=0}^{N}(x_{k+1}-x_k)^2 + {\alpha \over 3}
\sum_{k=0}^N(x_{k+1}-x_k)^3 , \nonumber
\end{eqnarray}
where $x_k$ is the $k$--th particle's displacement with respect to equilibrium
and $y_k$ its canonically conjugate momentum. Fixed boundary conditions
are set by $x_0=x_{N+1}=0$. 

In the original FPU framework, the FPU Hamiltonian 
is a first order perturbation of the $N$ coupled harmonic oscillators.
The Toda lattice, on the other hand, whose dynamics is described by the Hamiltonian function:
\begin{eqnarray}  \label{B1a}
H_{T}= \sum_{k=1}^{N} y_{k}^{2} + \frac {1}{4\alpha ^{2}}
\sum_{k=0}^{N} e ^{2\alpha (x_{k+1}-x_{k})}-\frac {N+1}{4\alpha
^{2}}  \nonumber
\end{eqnarray}
has a higher order tangency with the FPU's vector field. 

Under the canonical transformation:
 {
\begin{eqnarray}\label{lintra}
(x_k,y_k) &=&\sqrt{2\over N}\sum_{q=1}^{N-1} (Q_q,P_q)\sin\left({qk\pi\over
N}\right)\nonumber
\end{eqnarray}}
the integrals of motion of the linear system, i.e. the energies of the normal modes,
take the form: 
\begin{equation}\label{harmonicene}
E_q=  {1\over 2} (P_q^2+ \Omega_q ^2 Q_q^2),~~q=1,2,\ldots,N-1, \nonumber
\end{equation}
where $\Omega _q = 2 \sin( q \pi /N)$ are the harmonic frequencies and $Q_q$,$P_q$ are the modal positions and momenta, respectively.

 {To excite a packet of $m$ low frequency consecutive modes, one considers the initial condition: 
$$(Q_q(0),P_q(0)) =A_q (\sin \varphi _q,\cos \varphi _q) $$
where $\varphi _q$ are the initial phases and the amplitudes $A_q$ are non--zero only for the packet modes $q \in \{ 1,\ldots,m\}$.
This fraction of modes excited is denoted by $M=m/N$.}
Actually, in the present work only zero phases have been considered.
We refer the reader to \cite{Livi} for the role of coherent and random phases.

\section{Energy localization overview } \label{elo}


\begin{figure*}
\centering
\resizebox{1.\columnwidth}{!}{
\includegraphics{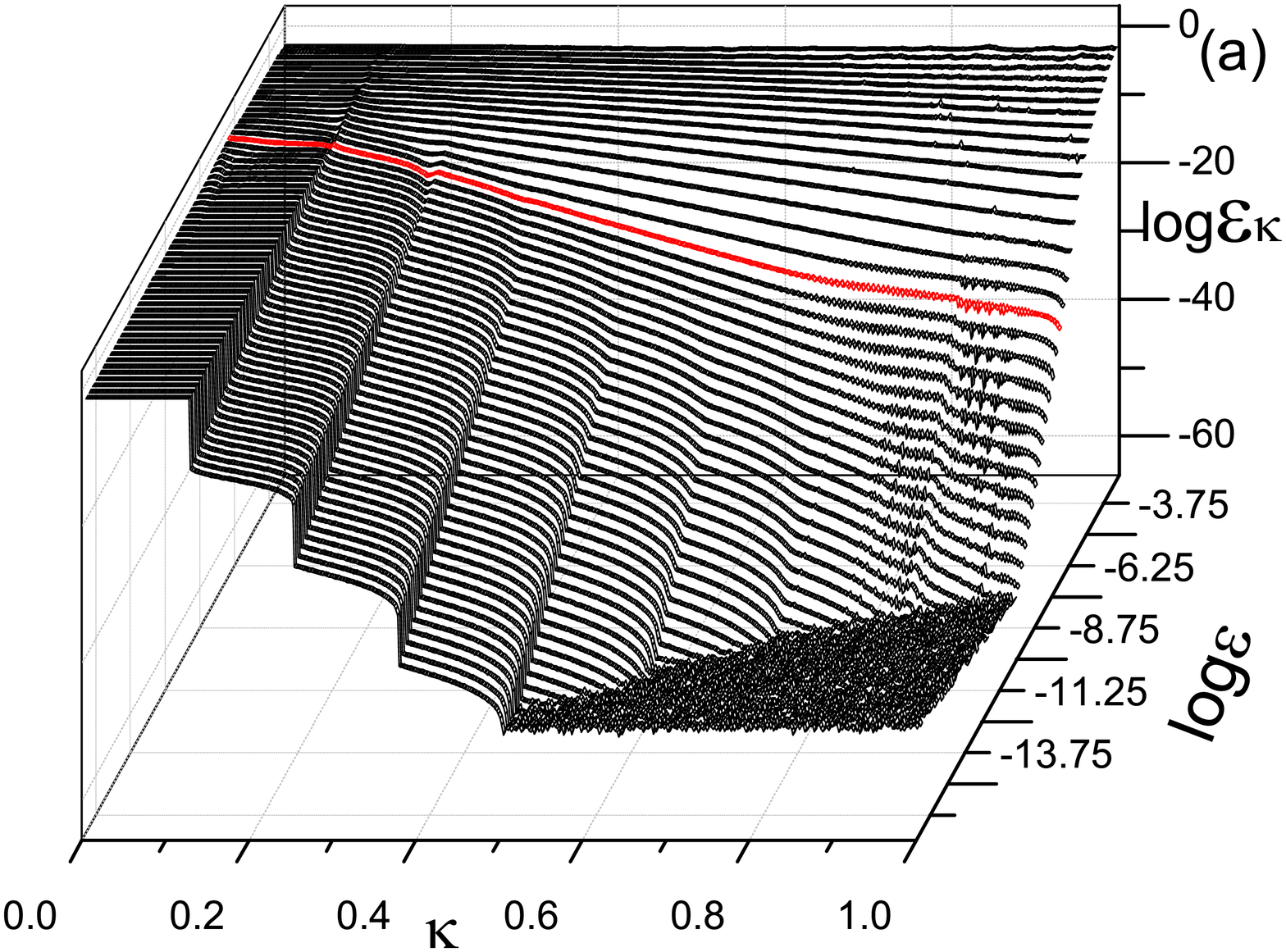} 
\includegraphics{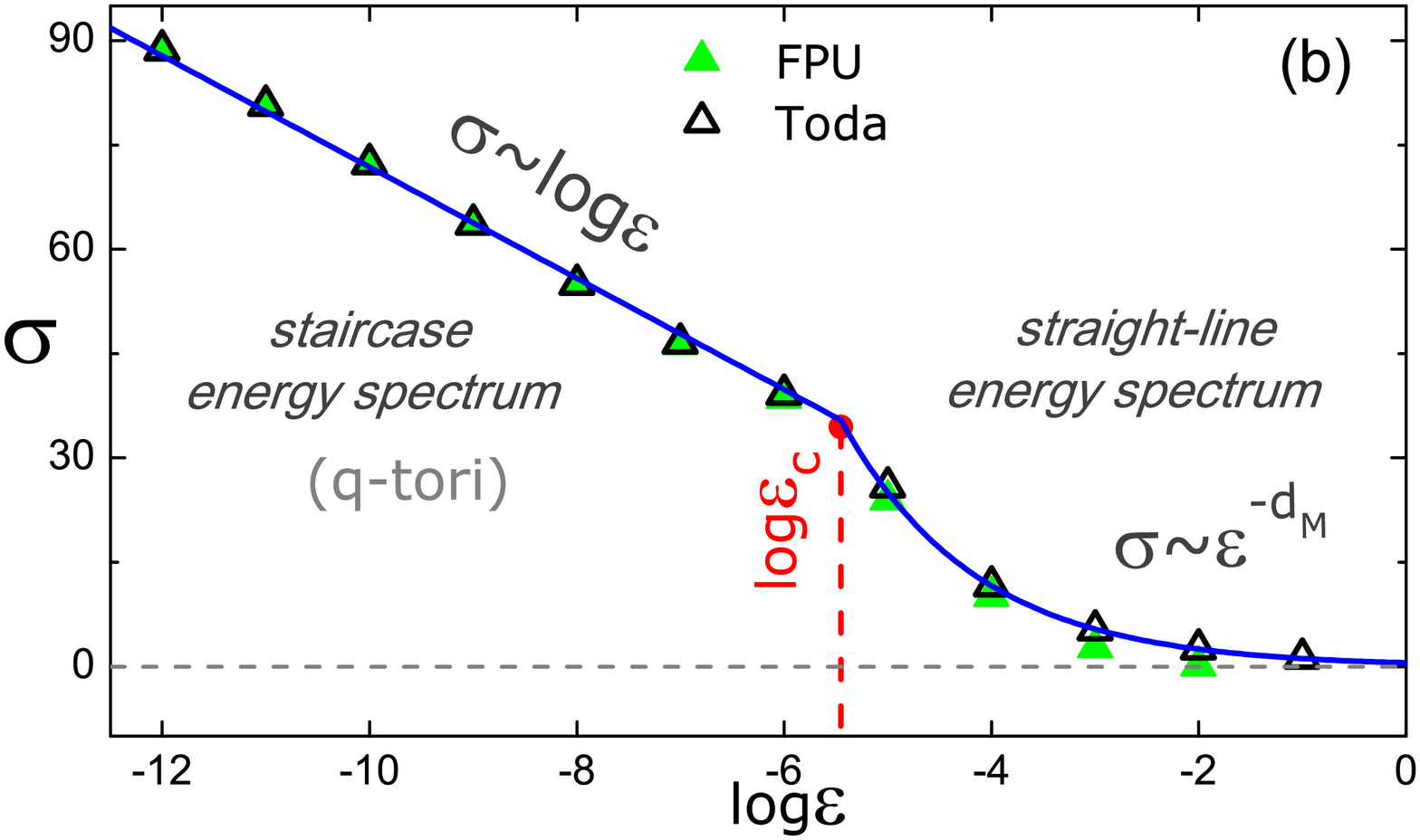}}
\caption{ (a)  {A bundle of Toda energy spectra, averaged over the time interval $[10^5,4\cdot 10^5]$, 
for $\alpha =0.5$, $M=12.5\%$ and $N=512$ at increasing energy values. Two localization patterns appear: 
a staircase below $\varepsilon_c=10^{-5.65}$ (with red) and a straight--line above.
(b) The slope of Toda (triangles), derived by a least--squares fitting to the energy spectrum of the panel (a),  
and the same for FPU (green triangles). The energy crossover $\varepsilon_c$ sharply} 
separates two localization laws.
\label{overview} }
\end{figure*}

Fig.\ref{overview}(a) shows the localization of the energy spectrum in  {Toda 
when a packet of $M=12.5\%$ modes} is initially excited at various energy levels. 
At very low energies the modal energy spectrum resembles a staircase.
This kind of profile is a particular characteristic of  $q$--tori and of trajectories lying close to them \cite{tori1,tori2}. 
However, as the total energy progressively increases, the division into groups disappears and the steps of modes 
merge in a straight--line profile.

The transition from the staircase exponential energy profile to the straight--line one 
 {is an {\it integrable transition} which occurs 
abruptly at the energy per particle value $\varepsilon_c$. A careful inspection on  
the slopes $\sigma$ of this example reveals the existence two different localization laws, 
which both appear in FPU and in Toda. In Fig.\ref{overview}(b) the slopes in both
systems are almost identical above and below $\varepsilon_c $. 
Nevertheless, in Toda the slope $\sigma$ remains constant in time, while
in FPU is expected to reach zero (energy equipartition) \cite{dresden,benettin1} in a second and longer time--scale.

Why two localization laws?}
When exciting packets of modes in FPU and in Toda at lower energies the 
trajectories lie on tori of dimension equal to the size of the seed packet, described by the   
frequencies $\omega _q\simeq \Omega _q$ \cite{tori1,tori2}, i.e. their frequencies are 
nearly their harmonic ones. Such solutions correspond to the energy profiles ($q$--tori and $q$--breathers case) 
which depend logarithmically on $\varepsilon$. At $\varepsilon=\varepsilon_c$ the first order
frequency corrections become significant and the phonon approach to the problem fails. 
We expect that above $\varepsilon_c $ there exists a set of non--commensurable Toda frequencies
which could be used to approximate the FPU frequencies and explain the further persistence
of energy localization. Such an approach could also prove the different localization 
law which is found numerically for $\varepsilon > \varepsilon _c$.

We denote the exponentially localized energy spectrum in FPU and in Toda by: 
\begin{equation}\label{enekappa}
\log {\cal{E}}_{\kappa }\simeq  - \sigma \kappa +\zeta , ~~~ {\scriptsize{0 \leq \kappa \leq 1}} , 
\end{equation}
where ${\cal{E}}_{\kappa }= E_{\kappa }/E$ are the normalized modal energies and $\kappa =q/N$ is the normalized wavenumber.
It is convenient to normalize the energy spectrum in order to compare the profiles of different energy
levels and system sizes. When exciting packets of modes and comparing the results for increasing $N$ values, the relevant way 
is to excite a fixed proportion $M$ of modes. In this way we excite modes up to a fixed frequency.

In the sequel our focus is to determine the expressions of the two localization laws for each energy regime. 




\section{At low energies: A staircase energy spectrum} \label{le}
\begin{figure*}
\centering
\resizebox{1.\columnwidth}{!}{
\includegraphics{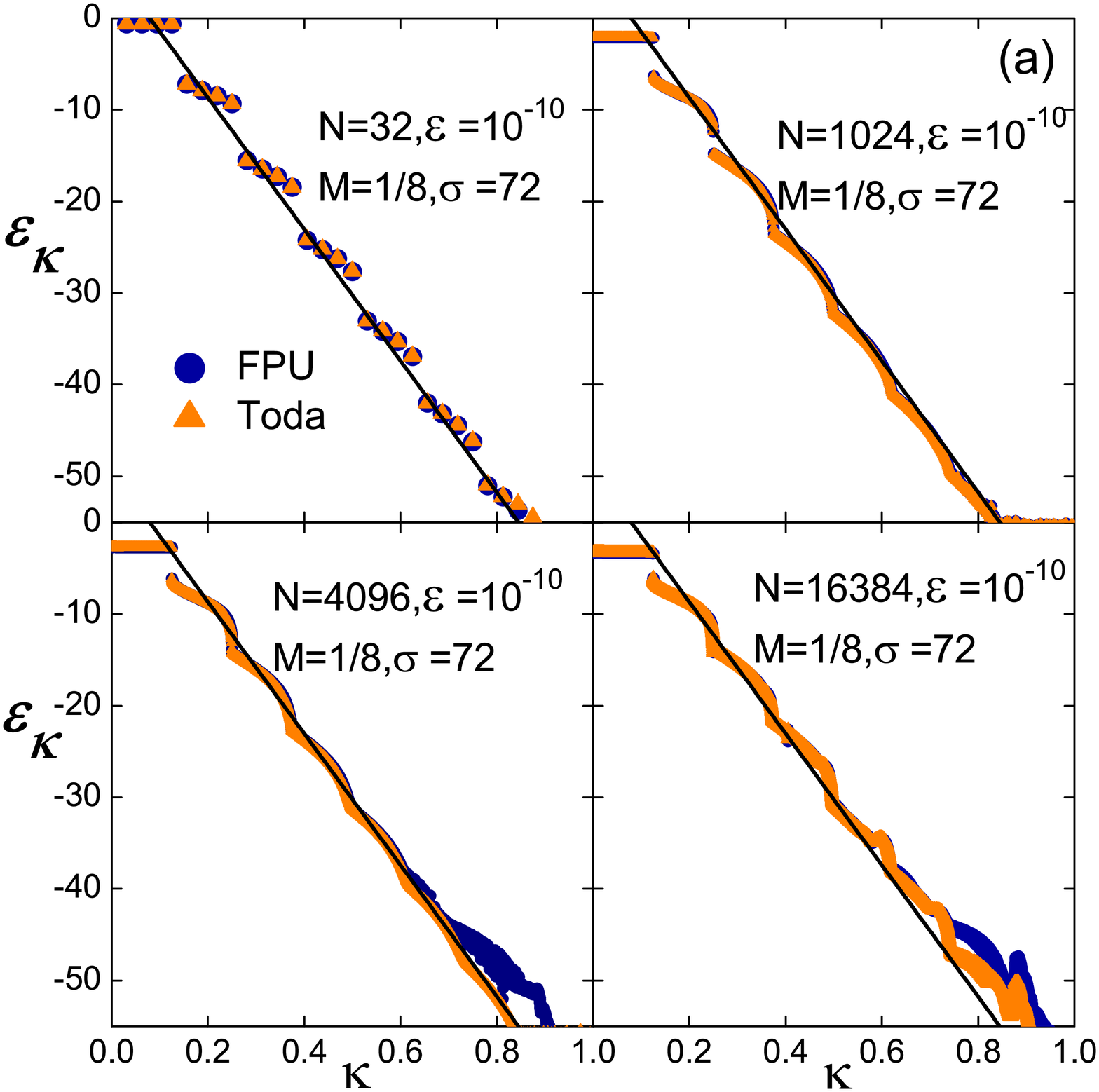} 
\includegraphics{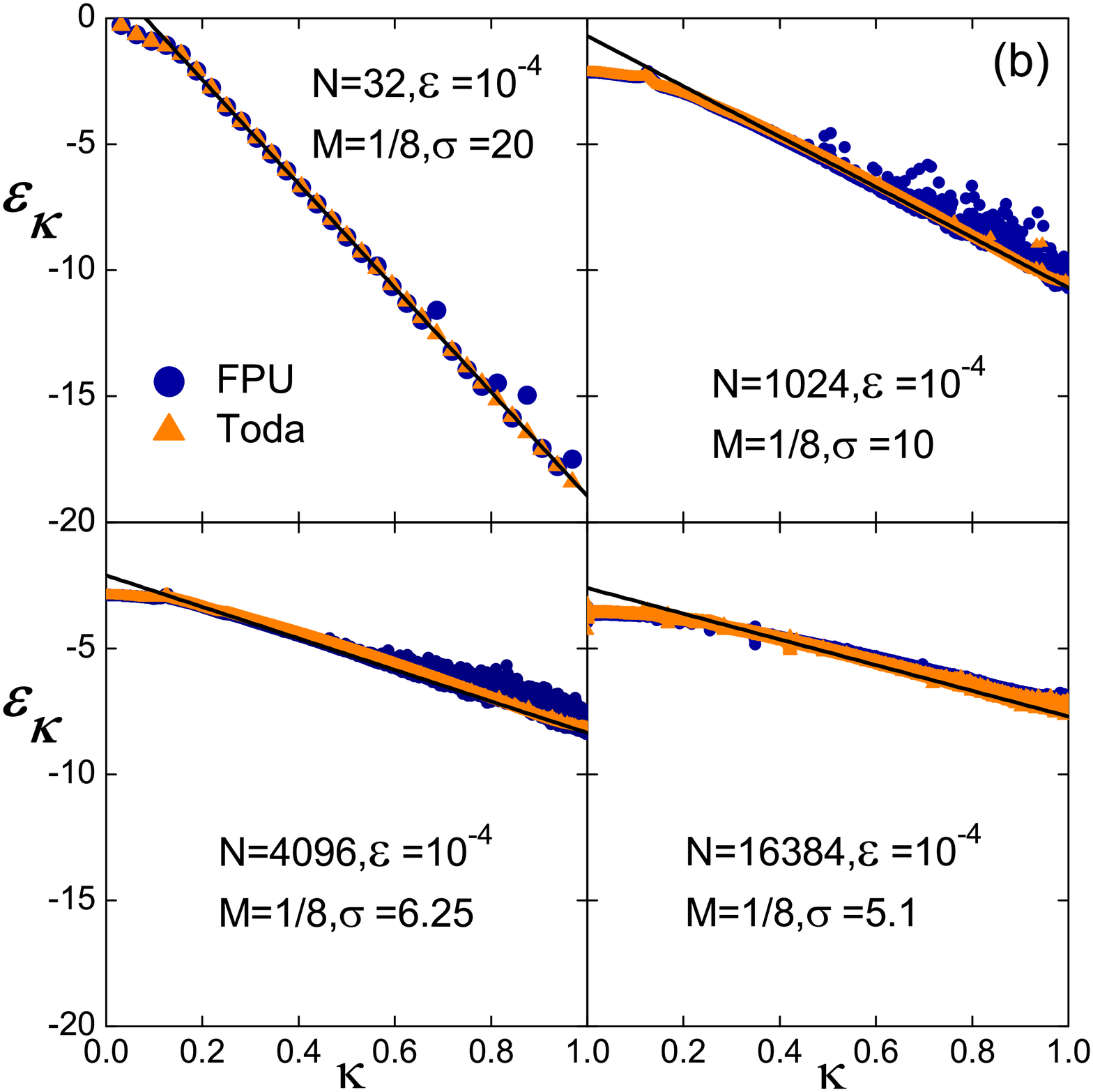}}
\caption{ The energy spectra of FPU (blue spheres) 
and Toda (orange triangles)  {averaged over the time interval $[10^5,4\cdot 10^5]$ and}
 for increasing, in powers of 2, values of $N$ for: 
(a) $\varepsilon <\varepsilon _c$ and (b)  $\varepsilon >\varepsilon _c$. \label{spectra}  }
\end{figure*}

Refs. \cite{tori1,tori2} reported the existence of $q$--tori in FPU which correspond to 
a near--continuation of packets of linear modes and result to exponentially localized energy profiles.
The energy propagation from the initially excited packet to the rest modes 
is found to cascade in exponentially decaying steps of groups of modes. 
In particular, the `seed modes' $\{ 1,\ldots,m\}$  result to the excitation of their $m$ consecutive modes 
$\{ m+1,\ldots,2m\}$, then to the modes $\{ 2m+1,\ldots,3m\}$ and so on. 
The localization law, i.e. the slope $\sigma$ in (\ref{enekappa}), for solutions on $q$--tori 
has been found to depend logarithmically on the energy density, as: 
\begin{equation}\label{slopelog}
\sigma = -  \frac{\log  \varepsilon}{M}  +\rho ,
\end{equation} 
where  $\rho = - M^{-1} \log ( {\alpha^2 }/{\pi^4 M^4})  $ and $M=m/N$ 
is the percentage of the consecutive lowest frequency modes 
which are initially excited.

In Fig.\ref{spectra}(a) we display the energy profiles of FPU and Toda for 
$M=12.5 \%$, $\varepsilon =10^{-10}$ and for 4 different $N$ values. Extensivity follows by the slope's $N$--independence:
$\sigma $ is constant in the panels of  Fig.\ref{spectra}(a) as long as $M$ and $\varepsilon $ 
have a constant value.

Despite the good superposition between
the two systems, non--integrability in FPU is revealed by the slightly detached high--frequency modes.
Moreover, the comparison to the Toda energy profile is a good method to detect the actual energy diffusion
in FPU and be assured that it is not due to numerical round--off errors.

\section{At high energies: A straight--line energy spectrum} \label{strlin}
\begin{figure*}
\centering
\resizebox{0.6\columnwidth}{!}{
\includegraphics{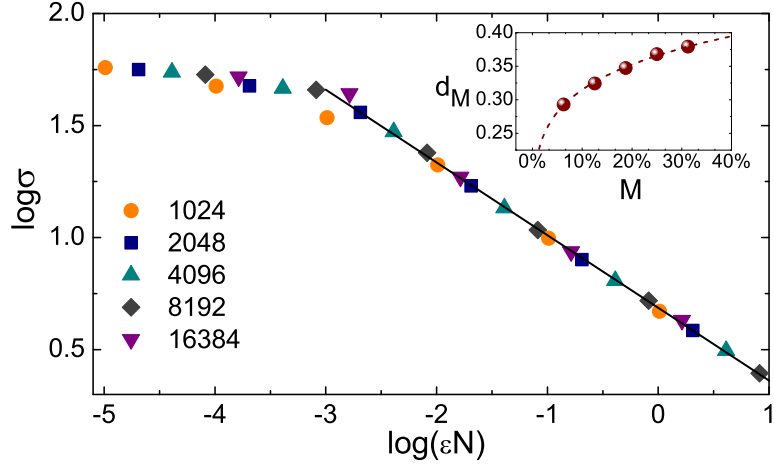} 
}
\caption{  The slope of the exponential Toda energy profile for $M=12.5 \%$ 
decays with $ \varepsilon N $ as $\sigma \sim (\varepsilon N) ^{-d_M}$. 
The fitting line is $-0.324\log(\varepsilon N) +0.687$. 
Inset: The numerical values for $d_M$ as a function of the packet size $M$. \label{2ndlaw}  }
\end{figure*}
A different localization law follows the energy spectrum of FPU and of Toda
at higher energies. The energy steps merge into a straight--line profile, 
characterized by a slope that tends to zero as $N\rightarrow \infty $.
The $N$--dependence of $\sigma $ is evident in the 4 examples of 
Fig.\ref{spectra}(b).

By systematically studying the exponential profiles in the Toda 
system (and not of FPU to avoid non--integrable tail fluctuations), we numerically derive
that the spectrum's slope takes the form: 
\begin{eqnarray}\label{sigma}
\sigma = \frac{C_M}{( \alpha^2 \varepsilon N) ^{d_M}}
\end{eqnarray}
where $d_M\simeq 0.458M^{0.163}$ (inset of Fig.\ref{2ndlaw}) and $C_M\simeq \pi  e^{-M/0.065} +2.63$. 
In particular, we arrived at the expression (\ref{sigma}) by observing that $\sigma$ scales 
with the energy $E= \varepsilon N$ for fixed $M \in [6\%,32\%]$ (as displayed in Fig.\ref{2ndlaw}). 
In agreement with Ref.\cite{Livi}, we find that the energy $E$ is the ruling parameter 
in the expression (\ref{sigma}) and this fact is attributed to the choice of {\it coherent phases}. 
Nevertheless, the case of random phases, which might make the system scale with $\varepsilon $ instead
of $E$, has been deferred to a future study. Furthermore, it is worth mentioning that the expression (\ref{sigma})
fails in the limit $M\rightarrow 0$. 
The slope for single--mode excitations follows the $N$--independent relation $\sigma =c(\alpha^2 \varepsilon )^{-1/4}$, 
derived by using a resonant normal form in the Korteweg--de Vries equation \cite{Ponno1,Ponno2}.

\begin{figure*}
\centering
\resizebox{0.6\columnwidth}{!}{
\includegraphics{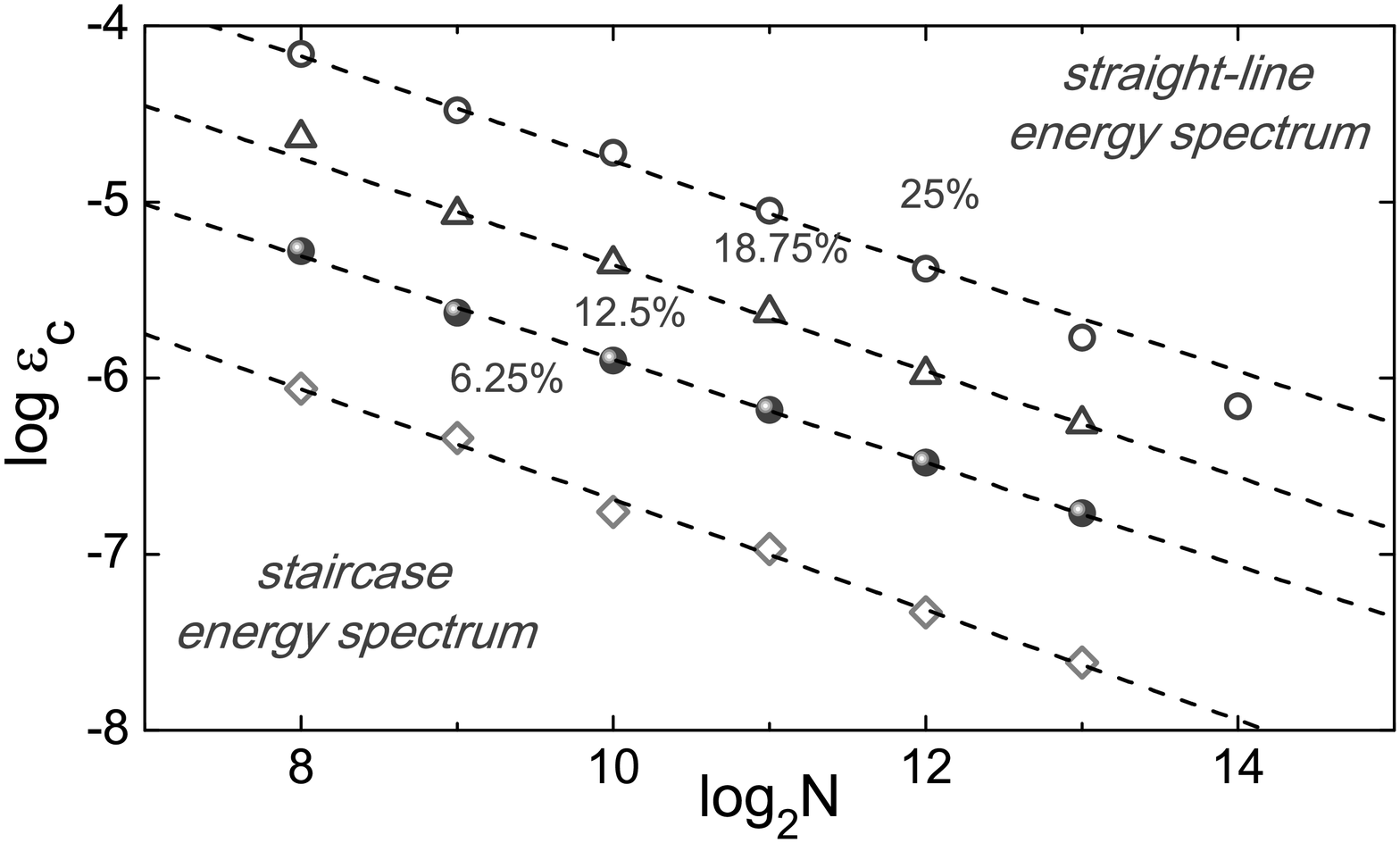} 
}
\caption{  The energy boundaries between the two regimes for the cases of $M=6.25 \%, 12.5 \%, 18,75\%$
and $25\%$  \label{crossover}    }
\end{figure*}

\section{In the thermodynamic limit} \label{regimes2}
We estimate the behavior of the energy crossover $\varepsilon_c$ as $N$ approaches infinity. 
We systematically studied the 4 cases: $M=6.25 \%$, $12.5 \%$, $18.75 \%$ and $25 \%$
for increasing system sizes of the Toda system. The results shown in Fig.\ref{crossover} suggest
that the energy crossover decreases with $N$ like $\varepsilon_c \sim 1/N$ and therefore any
kind of $q$--tori will disappear in the thermodynamic limit.

A semi--analytical estimation of $\varepsilon _c$ comes from the intersection of the two 
slope expressions (\ref{slopelog}) and (\ref{sigma}), which is 
$y^{d_M} \log y = -C_M M / (\pi^4  M^4 N)^{d_M}$, for  $y= \alpha^2 \varepsilon_c \pi ^{-4} M^{-4}$. 
In the limit $y\rightarrow 0$, the term $y^{d_M} \log y$ is well approximated by $-y^{d_M}/d_M$ and
therefore in the large $N$ limit we estimate the energy crossover:
$$\varepsilon_c \sim \frac{(d_M C_M M)^{1/d_M}}{\alpha^2 N}.$$

\section{Stickiness times in coherent packet excitations}
\label{equip}
\begin{figure*}
\centering
\resizebox{0.75\columnwidth}{!}{
\includegraphics{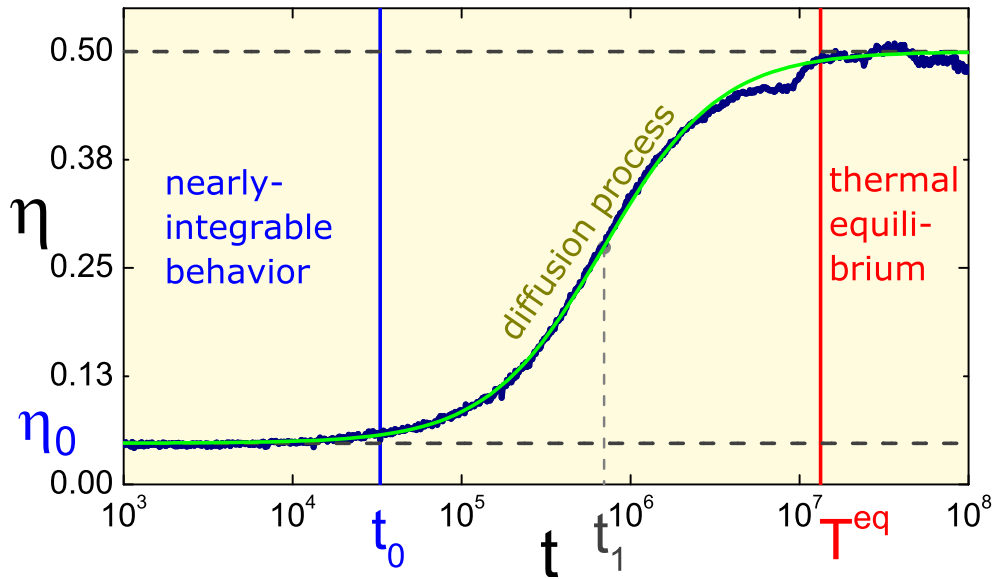} 
 }
\caption{  The temporal evolution of FPU's tail energy for $N=4096$, $M=12.5\%$ and $\varepsilon =10^{-2.5}$.
The fitting line is $-0.45\cdot 10^{(-t^{1.097}/7.5 \cdot 10^{6})}+1/2$.  \label{eta_tail}  }
\end{figure*}

 {In this section we focus on the stickiness, as well as on the equipartition times
in some cases, of extensive packet excitations. The crucial question is to
find the energy range in which the stability times extend exponentially and relate it with $\varepsilon_c\propto 1/N$.
In other words, how do these $q$--tori ($\varepsilon < \varepsilon_c$) and Toda--tori ($\varepsilon > \varepsilon_c$)
affect the stickiness times.
We partially answer to this problem for the reason that there are CPU--time limitations which make most of the 
computations below $\varepsilon_c$ unfeasible. Nevertheless, we stress that stickiness times greatly improve these CPU--times and 
are particularly suitable in detecting the variations of the actions\footnote{Namely, the variations of the Toda integrals in the FPU model.}.}

An accurate method to estimate  {both the stickiness and the equipartition times} in FPU is 
by monitoring the temporal evolution of the sum of the last half of the modes, defined as the {\it tail energy}: 
\begin{eqnarray}\label{eta}
\eta (t) =\sum_{\kappa =1/2}^{1} \varepsilon _{\kappa } (t). 
\end{eqnarray}
This method was introduced in \cite{dresden,benettin1}  {for calculating the equipartition times} 
and is very sensitive in detecting energy diffusion since high modes are more sensitive than the low ones 
to the non--integrability of the FPU model \cite{benettin2,flachlast}. 
In addition, this method produces very similar results when compared to more modern methods \cite{flachlast}.

A sketch of the tail energy evolution in time is illustrated in Fig.\ref{eta_tail}. 
 {There are three stages describing this evolution: (i) a nearly--integrable behavior
up to the stickiness time $t_0$, (ii) a sigmoidal diffusive process, and (iii) the final state of the 
system's thermal equilibrium for times $\geq T^{eq}$. Stickiness and equipartition times resemble 
the two sides of the same coin, namely, the departure from integrability and the arrival in equilibrium
of an FPU--trajectory.}

 {During the time where the tail energy is nearly constant $\eta_t \simeq  \eta_0$,
the FPU--trajectory stays close to its integrable counterpart, in the sense that $J_t \simeq J_0 $
where $J$ is a Toda integral \cite{dresden}. However, for times greater than $t_0$ the energy spreading among the modes results 
in a sigmoidal increase of the tail energy \cite{benettin1} (see Fig.\ref{eta_tail}), approximately described by: }
\begin{eqnarray}\label{eta}
\eta (t) = 1/2 + \frac{\eta_0 -1/2} { 1 + (t/t_1)^{\gamma } },~~~\gamma >0,
\end{eqnarray}
where $t_1$ corresponds to the `mid--graph's point' $\eta (t_1)= \eta_0/2 + 1/4 $. 
 {We note that the rate of $\eta (t)$ follows a bell--shaped distribution 
(logistic), with a maximum at $t=t_1$. In addition,} this sigmoidal evolution is a well--known diffusion process often encountered in population 
growth models \cite{Verhulst} and its universality is stressed in \cite{Scurve}.

\begin{figure}
\centering
\includegraphics[scale=0.20 ]{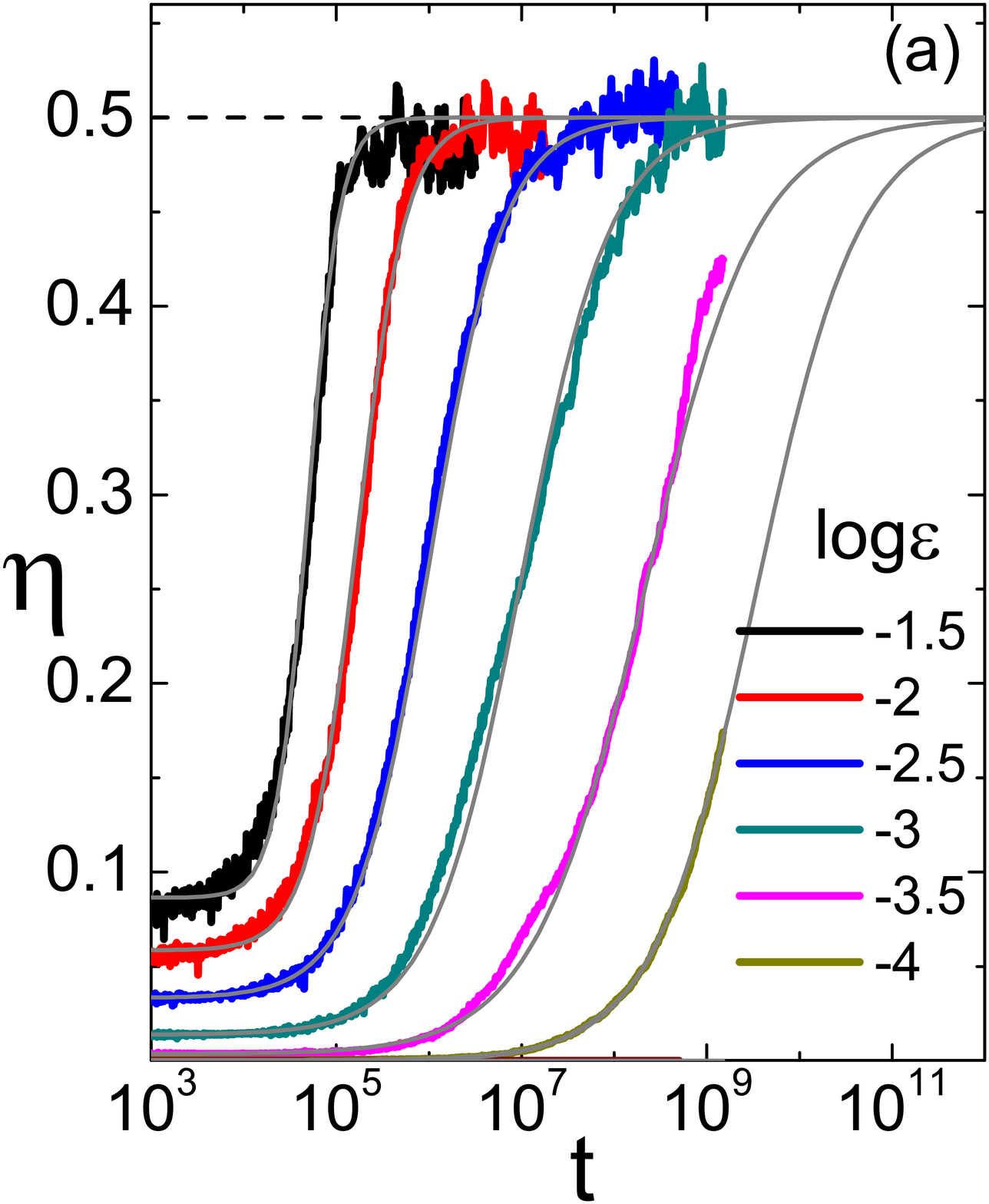} 
\includegraphics[scale=0.25 ]{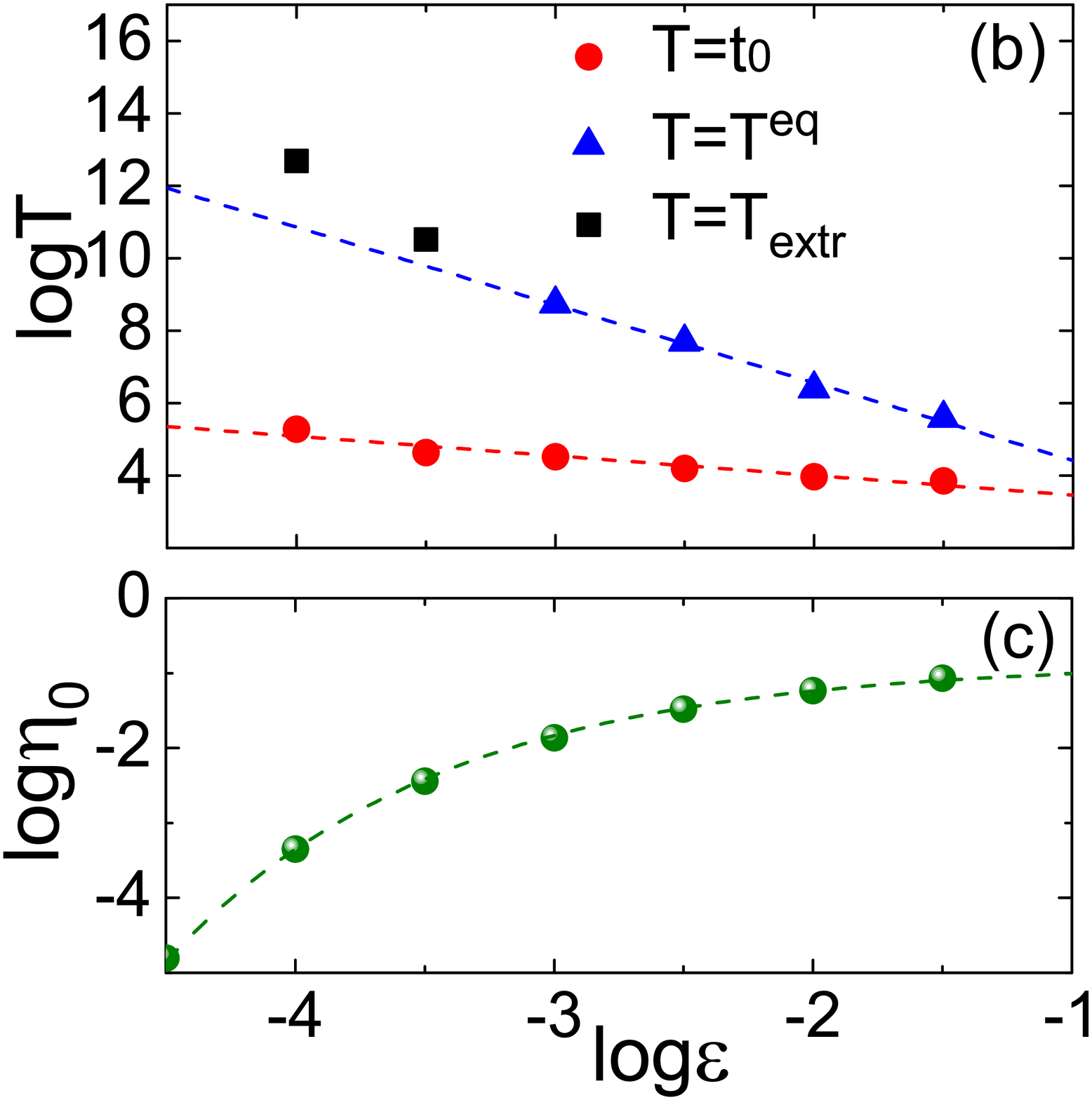}
\caption{ (a)  The temporal evolution of $\eta (t)$ for $N=2048$, $M=12.5\%$ and its fitting curve for decreasing energy values.
(b) The stickiness times with (red) circles fitted by  $\log t_0 \simeq  -0.55 \log \varepsilon +2.9$, 
the equipartition times with (blue) triangles fitted by  $T^{eq} \propto \varepsilon^{-2.15} $ for $\varepsilon > 10^{-3.5}$
and the extrapolated equipartition times with (black) squares for $\varepsilon \leq  10^{-3.5}$. 
(c)  The exponential decay of $\eta _0$ values with respect to $1/\varepsilon $, fitted by 
$\log \eta _0  \simeq -0.8623 -0.0573 \varepsilon ^{-0.41}$.  \label{eta_tail2}  }
\end{figure}

\begin{figure*}
\centering
\resizebox{0.75\columnwidth}{!}{
\includegraphics{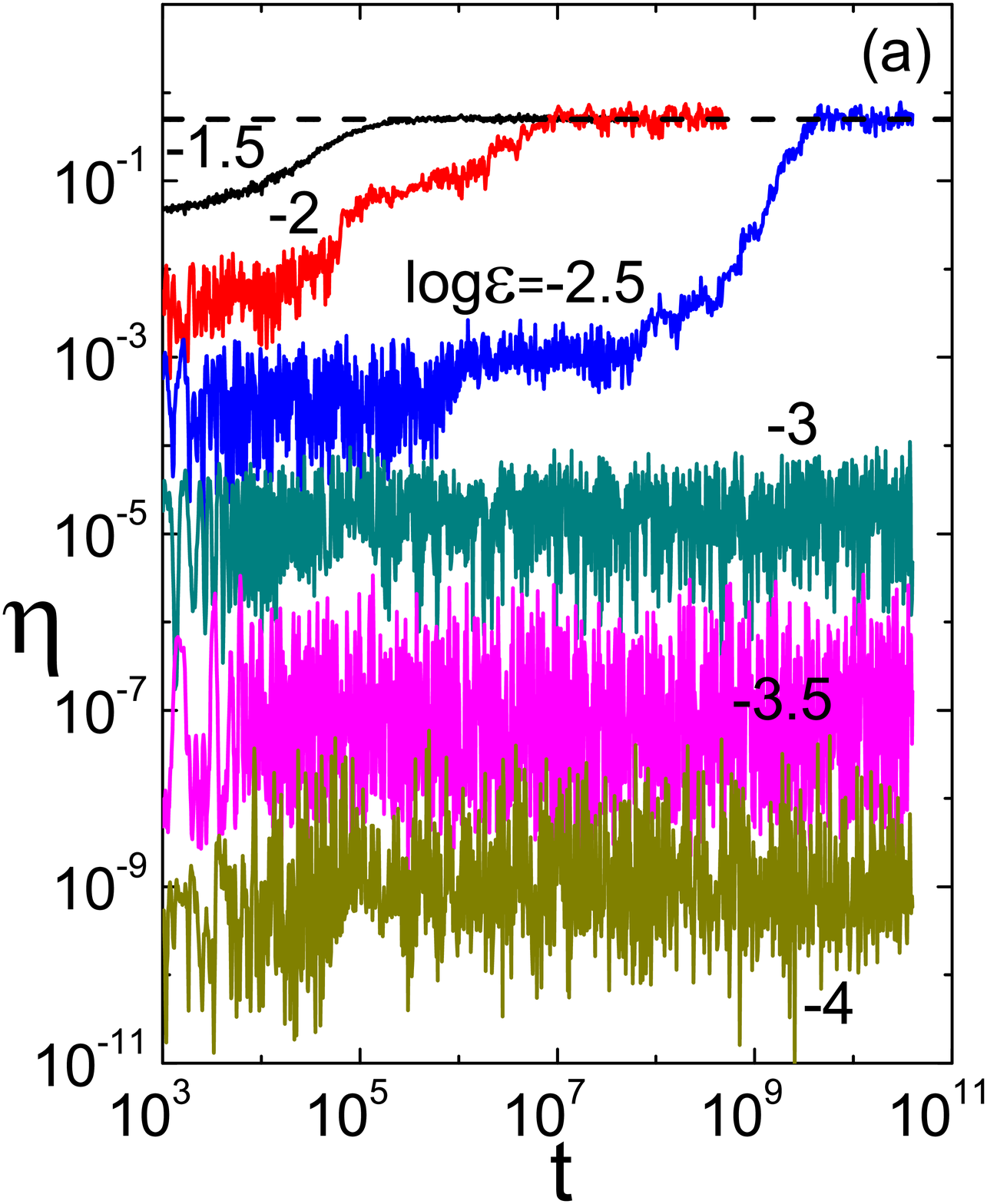} 
 \includegraphics{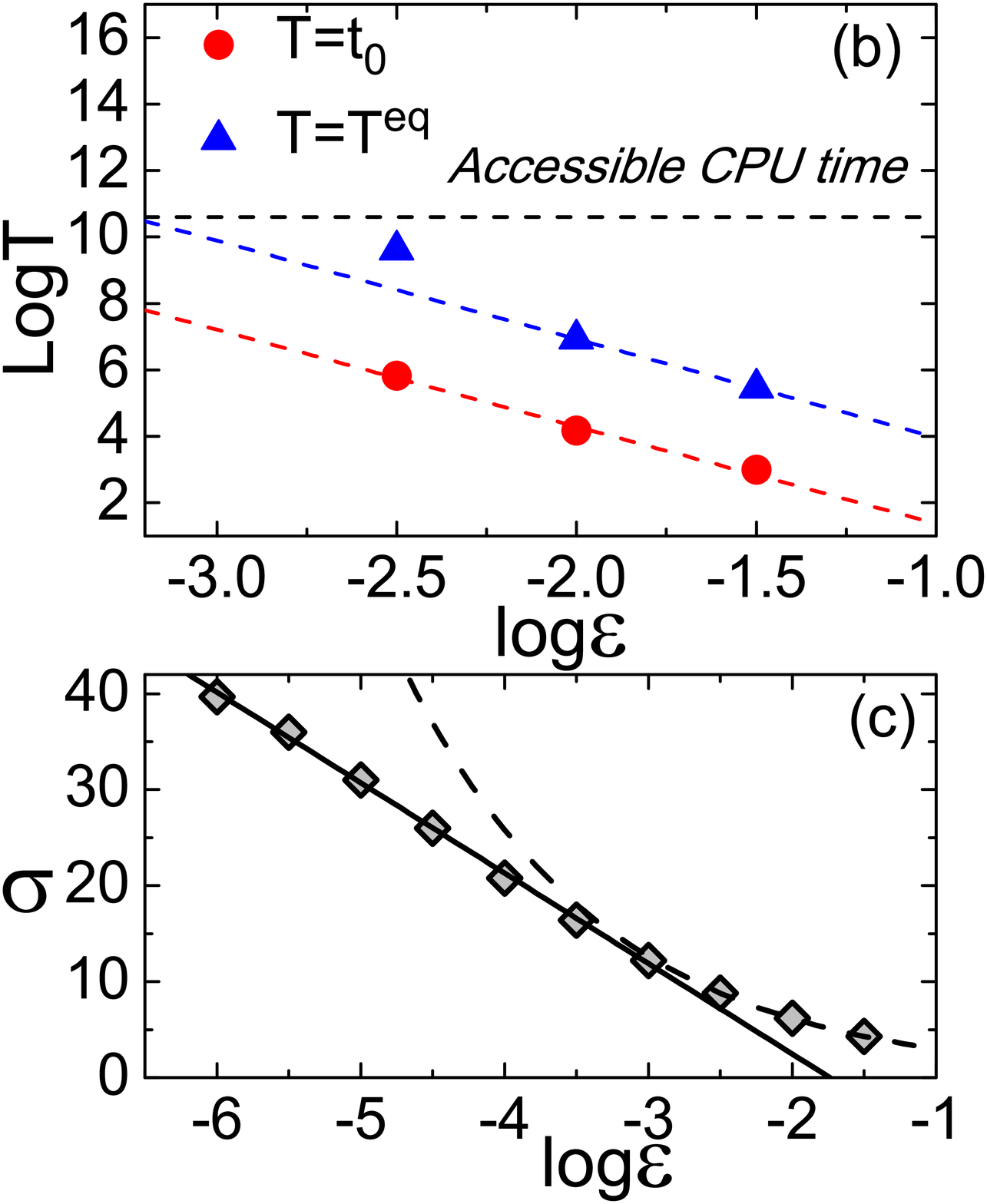}}
\caption{ { Panels (a),(b): Same as Fig.\ref{eta_tail2}(a),(b) for $N=32$, respectively.} 
(c) The slope of Toda's profile faces a transition at $\varepsilon _c =10^{-3}$. \label{last}   }
\end{figure*}

 {In the example of $N=2048$ particles with $M=12.5\%$ the critical energy is $\varepsilon _c=10^{-6.2}$.
We display the tail energy evolution of for a set of $\varepsilon >\varepsilon _c$ values in Fig.\ref{eta_tail2}(a)  
and report their corresponding equipartition and stickiness times in Fig.\ref{eta_tail2}(b).}
With (blue) triangles we denote the  {observed} equipartition times of  Fig.\ref{eta_tail2}(a), i.e. when the $S$--curved tail
has reached $1/2$, while with (black) squares the extrapolated equipartition times, predicted by the fitting line 
(\ref{eta}) on the incomplete data. It is not yet clear if below  {$\varepsilon =10^{-3.5}$, 
which is greater than $\varepsilon _c$, the extrapolated 
equipartition times  become a stretched exponential in terms of $1/\varepsilon$ 
and it is very hard to find $T^{eq} \sim  \mathcal{O}(10^{13})$ precisely. Moreover, finding empirically\footnote{To estimate  $\eta _0$ we need to know the expression (\ref{enekappa}). $\sigma $ in known from Eq.(\ref{sigma}) but not $\zeta $.} that the $\eta _0$--level decreases exponentially with $1/\varepsilon $ like $\eta _0 \sim \exp(-c/\varepsilon ^{0.41})$ (Fig.\ref{eta_tail2}(c)), it is yet possible that this decrease is responsible 
of the exponential stretch in the extrapolated equilibrium times. 
On the other hand,} the stickiness times clearly follow  a powerlaw 
$t_0 \propto  \varepsilon ^{-0.55}$, which implies that this energy range does not correspond to a Nekhoroshev regime,
where orbit--trapping is due to a dense set of KAM tori.

 {A systematic study on various $N$ values and $M$ packet sizes reveals that above
$\varepsilon _c$ stickiness times are indeed a power--law in terms of $1/\varepsilon$. The great challenge now is to repeat 
for the $q$--tori regime ($\varepsilon<\varepsilon _c$).}
Unfortunately, the case of $N=2048$ particles, which could give very clear results, is a notoriously difficult 
task that requires integration times greater than $10^{15}$ with {\it quadruple precision}. We therefore reduced significantly the system 
size to $N=32$ particles and paid the price of strong fluctuations and finite--size effects. 
 {The temporal evolution of the tail energies is reported in Fig.\ref{last}(a) and the times $t_0$, $T^{eq}$
in Fig.\ref{last}(b). Going beyond $t=10^{10}$ for energy values below $\varepsilon _c=10^{-3}$ (this crossover derives 
from Fig.\ref{last}(c)), we find that both the stickiness and the equipartition times become {\it at least} exponentially--long 
with respect to $1/\varepsilon $, a fact which  suggests that $\varepsilon _c$ is the boundary of a Nekhoroshev regime.

 \begin{figure*}
\centering
\resizebox{0.7\columnwidth}{!}{
\includegraphics{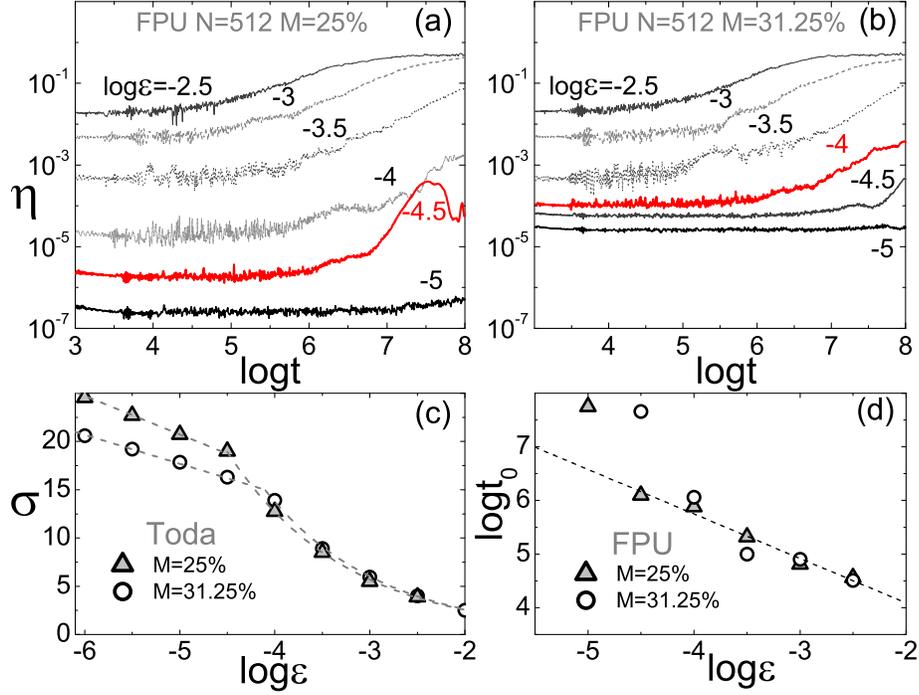} 
}
\caption{{The tail energy for each of the $512$--particle systems with (a) $M=25\%$ and (b) $M=31.25\%$ 
is displayed in the two upper panels. (c) The slope of the Toda profiles with $N=512$ 
and $M=25\%$ (triangles) and $M=31.25\%$ (circles) marks the energy crossovers 
$\varepsilon_c(25\%)=10^{-4.5}$ and $\varepsilon_c(31.25\%)=10^{-4.1}$. (d) Stickiness times detach 
from the dashed fitting line $t_0\propto \varepsilon ^{-0.83}$ below $\varepsilon_c(M)$ for each case. }\label{256_512}  }
\end{figure*}

We report two more examples supporting the same fact, namely $M=25\%$ and $M=31.25\%$ for $N=512$ particles in the Fig.\ref{256_512}.
Below $\varepsilon _c$ (for each case) the rising of the tail occurs in later than expected times, which 
can be naturally assumed to be the beginning of an exponential stretch. 
Nevertheless, we do not yet know if this is true for higher $N$ values, a statement
which would directly imply that the exponential--stability regime vanishes as $1/N$ in the thermodynamic limit.}

\section{Conclusions}
\label{concl}
In conclusion, we examined the existence of two 
well--separated energy regimes  {in terms of their energy localization properties, which appear in Toda and FPU
when coherent packet excitations are initially considered. In particular,   
the profile described by the slope $\sigma \sim \log (1/\varepsilon )$ in the phonon regime becomes
$\sigma \sim 1/(\varepsilon N)^d$ when the energy of the system is increased. The latter localization law, 
which is solely due to Toda, is the dominant one since the border separating them decays as $\varepsilon _c\propto 1/N$.
In this regard, a study on the FPU's stickiness times suggests that the exponentially--long stability times can be
found only below $\varepsilon _c$, a fact which implies ergodicity in the thermodynamic limit.
Based on these results, we find that an extension of the present study to a wider class of initial conditions
can better approach the resolution of the paradox.}

\paragraph{Acknowledgments}
We acknowledge very fruitful discussions with C. Efthymiopoulos. 
This research was supported by the State Scholarship Foundation (IKY) operational Program: `Education and Lifelong 
Learning--Supporting Postdoctoral Researchers' 2014-2020, and is co--financed by the European Union and 
Greek national funds.

\end{document}